\documentclass[aps,showpacs,superscriptaddress,twocolumn]{revtex4}
\usepackage{graphicx}
\begin{document}

\title{Volcanic forcing improves Atmosphere-Ocean Coupled General Circulation Model scaling performance}

\author{\surname{Dmitry} Vyushin}
\email{vjushin@ory.ph.biu.ac.il}
\affiliation{Minerva Center and Department of Physics, 
Bar-Ilan University, Ramat-Gan 52900, Israel}

\author{\surname{Igor} Zhidkov}
\email{zhidkov@shoshi.ph.biu.ac.il}
\affiliation{Minerva Center and Department of Physics, 
Bar-Ilan University, Ramat-Gan 52900, Israel}

\author{\surname{Shlomo} Havlin}
\email{havlin@ophir.ph.biu.ac.il}
\affiliation{Minerva Center and Department of Physics, 
Bar-Ilan University, Ramat-Gan 52900, Israel}

\author{\surname{Armin} Bunde}
\email{Armin.Bunde@theo.physik.uni-giessen.de}
\affiliation{Institut f\"ur Theoretische Physik~III, 
Justus-Liebig-Universit\"at Giessen,
Heinrich-Buff-Ring 16, 35392 Giessen, Germany}

\author{\surname{Stephen} Brenner}
\email{sbrenner@mail.biu.ac.il}
\affiliation{Department of Geography, Bar-Ilan University, Ramat-Gan 52900, Israel}

\date{\today}

\begin{abstract}
Recent Atmosphere-Ocean Coupled General Circulation Model (AOGCM) simulations of the twentieth century climate,
which account for  anthropogenic and natural forcings, make it possible to study
the origin of long-term  temperature correlations found in the observed records. We study 
ensemble experiments  performed with the NCAR PCM for 10 different historical
scenarios, including {\it no forcings,  greenhouse gas, sulfate aerosol,
ozone, solar, volcanic forcing} and various 
combinations, such as {\it natural, anthropogenic and all forcings.} 
We compare the scaling exponents characterizing the long-term
correlations  of the observed and simulated model data for 16 
representative land stations and 16 sites in the Atlantic Ocean for
these scenarios.  We find that inclusion of volcanic forcing in the AOGCM considerably improves
the PCM scaling behavior.  The scenarios containing volcanic forcing
are able to reproduce quite well the observed scaling  exponents for the land with
exponents around 0.65 independent  of the station distance from the
ocean. For the Atlantic Ocean, scenarios with the volcanic forcing
slightly  underestimate the observed persistence exhibiting an average 
exponent 0.74  instead of 0.85 for reconstructed data.  

\end{abstract}
\maketitle

\noindent

While many modelers and climatologists focus their studies on trends caused by
natural and anthropogenic forcings during the twentieth century~\cite{Stott,Mitchell,Meehl,Ammann,Santer,Storch}
we here focus on another important aspect of temperature anomalies - long-term correlations. 
One of the first studies on the question how general circulation models reproduce observed climate variability 
was performed by~\cite{Manabe}. Using power spectrum analysis, which 
is affected by nonstationarities in time series, they argue that the GFDL AOGCM reproduced 
the natural climate variability on decadal and centennial scales
correctly for a 1000 year control run integration. Several recent
studies of~\cite{KBunde1,KBunde2,Pelletier2,Talkner,Pelletier1,Caballero,Eichner}
clearly demonstrate that surface air temperature (SAT) anomalies are long-term correlated with a fluctuation 
exponent $\alpha$ close to 0.65. On the other hand, results
of~\cite{Bell,Govindan1,Govindan2,Syroka,Vjushin} indicate that AOGCMs 
underestimate surface air temperture (SAT) persistence for control run, greenhouse gas 
forcing only and greenhouse gas forcing plus sulfate aerosols scenarios. In contrast, 
recent studies of reconstructed data~\cite{Fraedrich,Blender}
claim that inner continental regions do not show long-term correlations, and thus AOGCMs successfully 
reproduce
 the natural persistence for the control run and greenhouse gas forcing only scenarios. 
 Recently, this claim was tested on observed records by~\cite{Bunde1} with the finding that the 
SAT fluctuation exponents for continental sites do show long-term correlation and the $\alpha$  values do not depend 
on the distance of the site from the ocean.

In this Letter we show that the recent PCM model simulations properly reproduce the observed 
long-term correlations for SAT on land only for  those scenarios that include 
volcanic forcing.  These scenarios also show better scaling
performance over the ocean than  the other scenarios.

In order to present the land-surface temperature profile 
for the last century, 16 observed daily maximum temperature time 
series are considered. They have been collected from different representative weather 
stations around the globe for the following sites: Vancouver, Tucson, Cheyenne, Luling, Brookings, Albany, Oxford, Prague, Kasan, Tashkent, Surgut, Chita, Seoul, Jakutsk,
Melbourne and Sydney.
We also analyze the gridded 
monthly mean sea surface temperature (SST) for 16 sites in the Atlantic ocean with a spatial resolution of 2.5$^o$x2.5$^o$ for the period 
of $1900-2002$ from the Kaplan Extended SSTA data set (see also~\cite{Monetti}). For $1900-1981$ this is the 
analysis of~\cite{Kaplan} which uses optimal estimation in the space of 80 
empirical orthogonal functions (EOFs) in order to interpolate ship observations of the 
U.K. Meteorological Office database~\cite{Parker}. The data after $1981$ 
consists of gridded data from the National Center for Environmental Prediction optimal 
interpolation analysis, which combines ship observations with remotely sensed data 
~\cite{Reynolds}. This analysis is performed  on the same set of 80 EOFs as used in 
~\cite{Kaplan} in order to provide enhanced data quality.


The model considered in our study is the Parallel Climate Model (PCM), 
which was developed at the National Center for Atmospheric Research (NCAR). It is a fully coupled
global ocean-atmosphere-sea ice-land surface model that produces a stable climate
without flux adjustment. The horizontal resolution of the atmosphere is equivalent to 
2.8$^o$x2.8$^o$, with 18 levels in the vertical. Resolution of the ocean is roughly
2/3$^o$, increasing to 1/2$^o$ at the equator, with 32 levels. The detailed description
of the model and results from experiments using various forcings and their combinations may be found in 
~\cite{Washington,Dai,Meehl} and~\cite{Ammann}. 

Here we study 10 forcing combinations:
{\it no forcings, greenhouse gas, sulfate aerosol, ozone, solar, volcanic,
solar + volcanic, ozone + solar + volcanic, greenhouse gas + sulfate aerosol + ozone, 
all forcings}. Greenhouse gas forcing is based on historical observations of CO$_2$, 
N$_2$O, CH$_4$, CFC-11, CFC-12, and ozone~\cite{Dai}.
Evolution of direct forcing from tropospheric sulfate aerosol is reported by 
~\cite{Kiehl}. Historical changes of solar irradiance were reconstructed
by~\cite{Hoyt} and volcanic forcing by~\cite{Ammann}.
The period of all experiments is 1890-1999.

For the no forcings and solar+volcanic scenarios we analyze the available 3-member ensembles,  
whereas for other scenarios 4-member ensembles are available. For each scenario, 
we selected the temperature records of the 4 grid points closest 
to each site, and bilinearly interpolated the data to the location of the observed site.

For each record, we analyse daily (or monthly) temperature anomalies $\Delta T_i$.
The $\Delta T_i$ are called long-term correlated if their autocorrelation function $C(s)$
decays with time lag $s$ by a power law 
\begin{equation}
\label{eq1}
C(s) \sim s^{-\gamma}, \qquad0<\gamma<1.
\end{equation}                                                 
To overcome possible nonstationarities in the data, we do not calculate $C(s)$ directly. Instead we 
construct the ``profile'' $Y_n = \sum_{i=1}^n \Delta T_i$ and study the fluctuation function $F(s)$ of the
profile in segments of length $s$ by using the second order detrended fluctuation analysis (DFA2)~\cite{Peng,Jan}. 
In DFA2 we determine in each segment 
the best second-order polynomial fit of the profile. The standard deviation of the 
profile from these polynomials represents the square of the fluctuations in each segment.

The fluctuation function $F(s)$ is the root mean square of the fluctuations in all 
segments. For the relevant case of long-term power-law correlations given by 
Eq.~(\ref{eq1}), with $0<\gamma<1$, the fluctuation function $F(s)$ increases 
according to a power law,
\begin{equation}
\label{eq2}
F(s) \sim s^\alpha, \qquad\alpha = 1-\frac{\gamma}{2}. 
\end{equation}
For uncorrelated data (as well as for short-range correlations represented by 
$\gamma \geq 1$ or exponentially decaying correlation functions), we have 
$\alpha = \frac{1}{2}$. For long-term correlations we have $\alpha > \frac{1}{2}$.

\begin{figure}
\noindent\includegraphics[width=20pc]{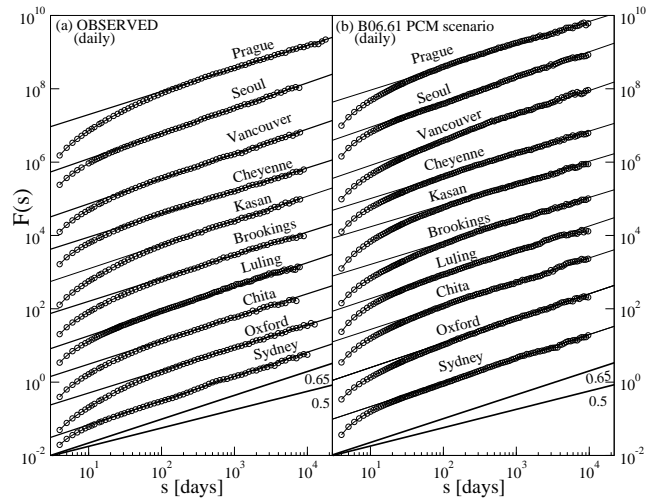}
\caption{ DFA2 fluctuation functions $F(s)$ for the daily surface air maximum temperature anomalies at 10 land sites:
 (a) observed data and  (b) NCAR PCM B06.61 (all forcings) simulated data. The scale of $F(s)$ 
is arbitrary. The straight lines crossing each curve represent the best asymptotic fit. 
The two lines shown at the bottom have slopes 0.65 and 0.5.}
\end{figure}

First we plot the results of DFA2 (DFA curves of higher order  show the same performance)
 for the observed daily maximum temperature (Figure 1a) and NCAR PCM simulations 
from the B06.61 run (Figure 1b). Run B06.61 represents one of the runs from the 
all forcings ensemble. All curves are shown in a double logarithmic presentation. 
We plot 10 typical DFA curves chosen from our 16 sites over land.
 The sites chosen include coastal, near coastal, and inland locations.

The approximate period of the observed records is 1880-1990, with the maximum length for 
Prague (1775-1992) and the minimum for Seoul (1908-1993). The period of the B06.61 run is 
1890-1999. The slopes in Figure 1a correspond to fluctuation exponents of the 
observed SAT anomalies, and vary from 0.62 to 0.68, with an average  close to 0.65. 
Figure 1a demonstrates that SAT anomalies for all sites studied 
obey long-term power-law correlations independent of the distance from the nearest 
ocean. The slopes in Figure 1b range from 0.62 to 0.69. Comparing Figures 1a and 1b shows that  the scaling of the NCAR PCM output  
agrees quite well  with the scaling of the observed data over land.

\begin{figure}
\noindent\includegraphics[width=20pc]{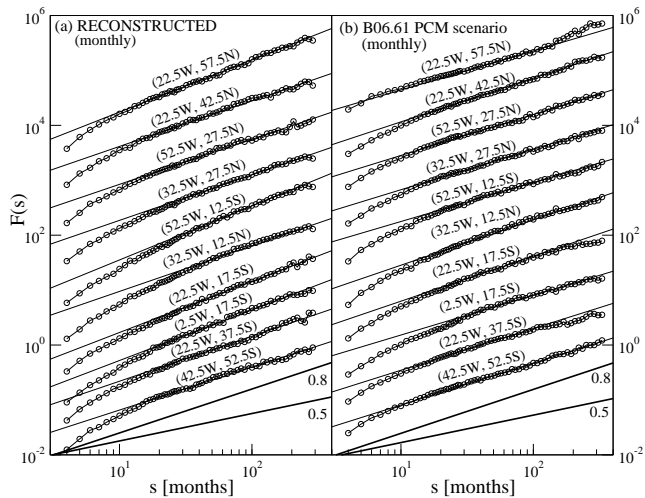}
\caption{ DFA2 fluctuation functions $F(s)$ for monthly sea surface temperature anomalies at 10 sites in the Atlantic ocean :
(a) the Kaplan reconstructed data and  (b) NCAR PCM B06.61 (all forcings) run.
The scale of $F(s)$ is arbitrary. The straight lines crossing each curve represent the 
best asymptotic fit. The two lines shown at the bottom have slopes 0.8 and 0.5.}
\end{figure}

Figure 2 shows DFA2 curves for the 10 sites in the Atlantic ocean for the Kaplan 
reconstructed monthly SST anomalies (Figure 2a) and for the NCAR PCM monthly averaged SST
anomalies from the all forcings B06.61 run (Figure 2b). The slopes for the reconstructed SSTA vary from 
0.71 in the equatorial part of the Atlantic to 1.0 in the Northern Atlantic, with an 
average of 0.85. 
The SSTA exponents characterizing  the memory effect on decadal and centennial 
scales seem to depend on complex ocean circulation dynamics. 
The variation of the scaling exponents over 
the Atlantic Ocean is significantly  larger than on land,  which is probably due to different 
ocean circulation patterns in equatorial, mid-latitude, and high-latitude regions. 
In a double logarithmic presentation, the slopes of the DFA2 curves for
the simulated ocean records have an average of 0.72 which is noticeably lower than the observed average of 0.85. 

\begin{figure}
\noindent\includegraphics[width=20pc]{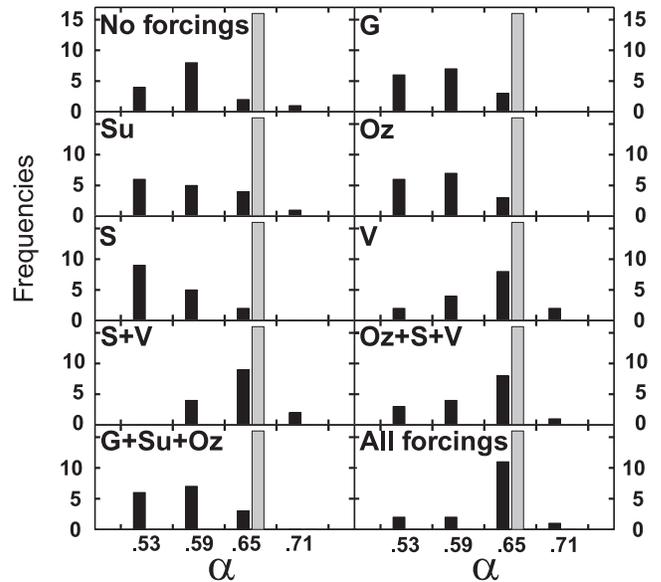}
\caption{ Histograms of the fluctuation exponents $\alpha$ for the observed records (grey column) and
the simulated records, for land stations. The considered 10 scenarios are: no
 forcings, greenhouse gas (G), sulfate (Su),  ozone (Oz), solar (S), volcanic (V), solar + volcanic (S+V), ozone +
 solar + volcanic (Oz+S+V), greenhouse gas + sulfate + ozone (G+Su+Oz), and all forcings. Four bins in each panel correspond to $\alpha$ 
 in the intervals [0.5,0.56), [0.56,0.62),[0.62,0.68), and [0.68,0.74] respectively. The grey column
in each panel corresponds to the fluctuation exponent distribution for the observed
records, the black columns are for simulated records.}
\end{figure}

Figure 3 presents the fluctuation exponent distribution for the observed
data and for the 10 NCAR PCM  scenarios considered for the 16 land locations. In each panel
the ``grey'' column in the range [0.62,0.68) represents the distribution of  the fluctuation exponents for the observed data.
As seen from the figure, scenarios containing volcanic forcing best reproduce the observations since they have a peak 
at the same range [0.62-0.68) as the observed data. Their average fluctuation exponent 
is close to 0.65.  In contrast, the other six scenarios that do not contain volcanic forcing, have an average 
fluctuation exponent for land less than 0.6 (see also~\cite{Govindan2}). 

Similar behavior is found over the Atlantic Ocean. Only those scenarios
containing the volcanic forcing exhibit an average fluctuation exponent greater
than 0.7 with the largest value equal to 0.76 for the volcanic forcing only scenario.
Figure 4 shows the fluctuation exponent distribution for the Atlantic Ocean for the Kaplan 
data (in grey) and for 10 studied scenarios (in black). Thus the PCM underestimates the fluctuation 
exponents obtained for reconstructed data
 for the Atlantic Ocean by 10-15\%. 

Therefore,  we can conclude that for the NCAR PCM addition of volcanic forcing to any other forcing 
combination immediately improves its scaling behavior both for the land and the ocean.
This fact suggests that (besides the atmosphere-ocean coupling) the volcanic forcing is  mostly responsible for the presence 
of the long-term correlations in the NCAR PCM over land on annual and decadal scales. 
For the ocean, the addition of volcanic forcing leads to stronger memory and consequently higher fluctuation exponents comparing 
to those for the land. However the NCAR PCM still underestimates the observed persistence of the oceans.

\begin{figure}
\noindent\includegraphics[width=20pc]{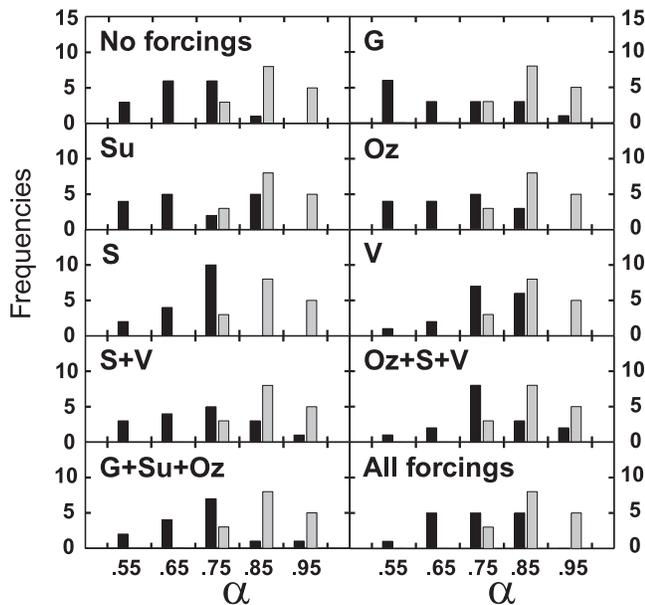}
\caption{Histograms for the fluctuation exponents for the reconstructed
(in grey) and simulated (in black) records for the Atlantic Ocean. The five bins
correspond to $\alpha$ values in the intervals [0.5,0.6),[0.6,0.7),[0.7,0.8),[0.8,0.9), and [0.9,1.0] 
respectively.}
\end{figure}

The main conclusion from our research is that the NCAR PCM is able to reproduce 
the scaling behavior of the observed land SAT records for 
the last century after taking into account all historically based natural and 
anthropogenic forcings. However, even the best scenario
for the land slightly underestimates natural SST persistence in the ocean, possibly due to errors in the  simulations of
deep ocean circulation, the atmosphere-ocean interaction, and/or an insufficiently long spin up period
 of
the ocean component of the AOGCM.

The results presented in this letter may  also help to clarify the controversy about the values 
of SAT fluctuation exponents for inner continental regions (see~\cite{Fraedrich}). 
Our study indicates that not only 
the observed records for inner continental regions are long-term correlated in agreement with
~\cite{Bunde1}, but also the recent PCM simulations for these regions show  similar fluctuation 
exponents,  characteristic of long-term persistence.

 Finally, this letter also supports the suggestion of
 ~\cite{Govindan2} that the inability of the seven leading
 AOGCMs, for their  control runs, greenhouse gas forcing only, and greenhouse gas plus 
aerosols scenarios, to mimic the observed SAT persistence is caused by the absence of 
natural forcings, in particular volcanic forcing. The results of  our detrended fluctuation analysis for the 16 land stations 
around the globe and the 16 sites in the Atlantic ocean 
suggest that volcanic forcing has by far the largest impact on the AOGCM long-memory persistence.

\begin{acknowledgments}
This work has been supported by the Deutsche Forschungsgemeinschaft and 
the Israel Science Foundation. We are grateful to W. M. Washington, M. Wehner, and G. Strand 
for providing access to the NCAR PCM simulations data. 
The authors thank J. Eichner, J. Kantelhardt and E. Koscielny-Bunde for fruitful discussions.

\end{acknowledgments}

\end{document}